\documentstyle[prl,aps,epsfig,citesort,amsmath]{revtex}

\def\vec#1{\boldsymbol{#1}}
\def\nn{\hat {\vec{n}}}
\def\bv{\vec{v}}
\def\br{\vec{r}}

\begin{document}
\twocolumn[\hsize\textwidth\columnwidth\hsize\csname
@twocolumnfalse\endcsname
\title{A global equation of state of two-dimensional hard sphere systems}
\author{Stefan Luding} 
\address{
Institute for Computer Applications 1, 
Pfaffenwaldring 27, 70569 Stuttgart, Germany }

\maketitle

\begin{abstract}
Hard sphere systems in two dimensions are examined for arbitrary
density. Simulation results are compared to the theoretical predictions
for both the low and the high density limit, where the system
is either disordered or ordered, respectively. The pressure
in the system increases with the density, except for an intermediate
range of volume fractions $0.65 \le \nu \le 0.75$,
where a disorder-order phase transition occurs.
The proposed {\em global equation of state} (which
describes the pressure {\em for all densities}) is applied to
the situation of an extremely dense hard sphere gas in a gravitational
field and shows reasonable agreement with both experimental and 
numerical data.\\

PACS number(s): 51.30.+i, 51.10.+y, 64.70.Dv, 45.70.-n
\vspace{-0.6cm}~\\
\end{abstract}
\narrowtext
\vskip1pc]

\def\FIGA{
\begin{figure}[htb]
\begin{center}
     ~\vspace{-0.9cm}\\
\epsfig{file=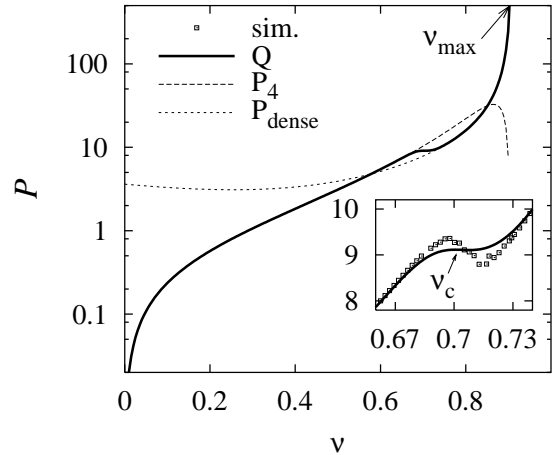,height=8cm,angle=-90}
\end{center}
\caption{Global, dimensionless equation of state $Q$, plotted against the
volume fraction $\nu$ with logarithmic vertical axis. The dashed and dotted
lines correspond to $P_4$ and $P_{\rm dense}$, respectively, see Eqs.\
(\ref{eq:defP}), (\ref{eq:g2a}), and (\ref{eq:pdense}).  In the
inset, simulation data ($N=1628$, $r=1$) are compared with $Q$.
}
\label{fig:figQ}
\end{figure}
}

\def\FIGB{
\begin{figure}[b]
  \center{
     ~\vspace{-0.8cm}\\
     ~\epsfig{file=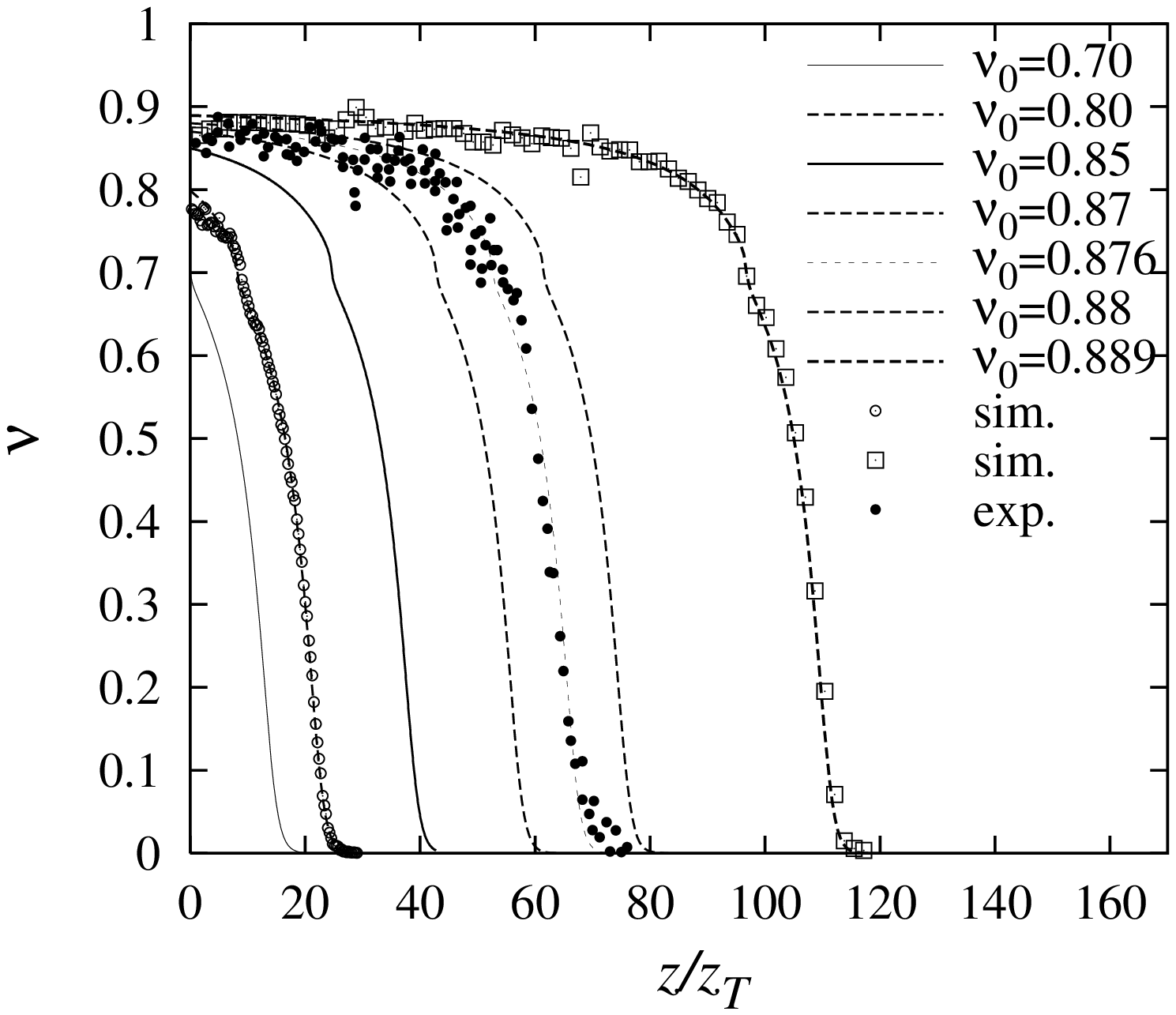,height=8cm,angle=-90,clip=}
     ~\vspace{0.5cm}\\
    }
  \caption{Volume fraction of the hard sphere gas as a
    function of the rescaled height $z/z_T$ for different
    $\nu_0$ values as given in the inset.  Lines are the
    theoretical predictions with increasing bottom density,
    from left to right, open symbols are two simulations
    and the solid dots are the experimental data from
    Ref.\ \protect\cite{clement91}.
    }
  \label{fig:grav1}
\end{figure}
}

A hard sphere (HS) system is a simple and tractable model for various 
physical phenomena.  It was used to examine disorder-order 
transitions, the glass transition, or simple gases and liquids 
\cite{ziman79,chapman60,hansen86,landau86,rascon96doliwa98} 
both theoretically and numerically.
The theory that describes the behavior of rather dilute hard
sphere systems is the kinetic theory \cite{chapman60,hansen86}, 
where particles are assumed to be rigid and collisions take 
place in zero time.  
An extension to Boltzmann's low density theory was 
introduced by Enskog \cite{chapman60,hansen86}, taking into account
excluded volume effects and also momentum transport via collisions. 
In the limit of high densities, the cage-effect,
where particles are captured by their neighbors \cite{rascon96doliwa98}, 
becomes important and a free-volume theory can be formulated
\cite{kirkwood50buehler51wood52,alder55alder59alder62}.
There exists no theory, to our knowledge, which is valid for
the intermediate densities where the system changes from the 
disordered to the ordered state, however, various theoretical approaches 
were proposed in the last decades, see Refs.\ 
\cite{torquato95velasco97,rascon96doliwa98,santos99ric,lowen00} 
and references therein.  

When dissipation is added to the HS model, one has the simplest version 
of a granular gas, i.e.~the inelastic hard sphere (IHS) model.
Granular media represent the more general class of dissipative, 
non-equilibrium, multi-particle systems \cite{behringer97,herrmann98}.  
Attempts to describe granular media by means of the kinetic theory
are usually restricted to certain limits like small densities
or weak dissipation \cite{jenkins85b,sunthar99}. 
Also in the case of granular media, one has to apply higher order 
corrections to successfully describe the system under more general 
conditions \cite{brey98,montanero99,sela98,noije98c} or for
multi-particle contacts \cite{multicont}.  The result
of a kinetic theory approach is, in the simplest case, a set of 
balance equations for mass, momentum, and energy with constitutive
expressions for the transport coefficients, describing stress, viscosity,
heat conduction, and energy dissipation.  In this letter, we focus
on the pressure $p$, the isotropic part of the stress in a hard sphere gas -- 
or in a granular gas in its elastic limit $r \rightarrow 1$, with the 
coefficient of restitution $r$. 
The model system is periodic and two-dimensional (2D) with volume 
$V=l_x l_y$, where $l_x$ and $l_y$ are horizontal and vertical size, 
respectively. It contains $N$
particles with radii $a$, and masses $m$ which are located at positions $\br_i$
with velocities $\bv_i$.  The fraction of the area which is covered by 
particles is denoted as volume fraction $\nu=N \pi a^2 / V$.
The kinetic energy is $E = \frac{m}{2} \sum_{i=1}^N \bv_i^2$,
the temperature is defined as $T=E/N$ in 2D and the energy density is $E/V$.
For low and intermediate densities $\nu < \nu_c$ (with the ``crystallisation''
density $\nu_c$ at which order becomes important), the kinetic theory leads 
to an expression for the equation of state, i.e.~the dimensionless excess
pressure due to particle interactions
\begin{equation}
P:=pV/E-1 = 2 \nu g(\nu) ~.
\label{eq:defP}
\end{equation}
For an ideal gas with non-interacting particles, one has $pV/E=1$ so that 
$P=0$; for non-zero densities one has $P>0$ since the collisions 
contribute to the momentum transport and thus to the pressure,
the viscosity, the heat-conductivity and the dissipation rate.
In cases with $r<1$ the factor 2 can be replaced
by $1+r$. The pair correlation function at contact $g(\nu)$
accounts for the probability that a collision occurs.  Typically,
$g(\nu)$ is determined via a virial expansion around low densities
and one can use
\begin{equation}
g_4(\nu) = \frac{1-7\nu/16}{(1-\nu)^2} -
              \frac{\nu^3/16}{8 (1-\nu)^4}  ~,
\label{eq:g2a}
\end{equation}
where the subscript $4$ indicates that the second term is of order
$1/(1-\nu)^4$. The first term in Eq.\ (\ref{eq:g2a}) is the simpler, 
frequently used version $g_2(\nu)$ introduced by Henderson 
\cite{henderson75henderson77,verlet82,jenkins85b}.
Note that the expression in Eq.\ (\ref{eq:g2a}) is slightly
different from the form in Refs.\ \cite{verlet82,sunthar99}.
The value of $g_{4}$, taken at contact, accounts for the excluded volume
effect and the increase of the collision rate with density.
At densities larger than $\nu_c \approx 0.7$, an ordered 
triangular structure is evidenced \cite{tobochnik88,luding00,torquato00}.

One of the unsolved problems concerning an application of the 
balance equations to a specific boundary value problem is 
the limited range of validity of Eq.\ (\ref{eq:g2a}).  Under
realistic conditions with $r<1$, the volume fraction can take values 
$\nu > \nu_c$ \cite{luding98f} so that a solution based on 
Eqs.\ (\ref{eq:defP}) and (\ref{eq:g2a}) can be correct 
up to $\nu_c$ only.  This is even worse, since the ``virial''
$\nu g(\nu)$ also occurs in all the other transport coefficients. 
For the same reason, fortunately, a generalization of the pressure 
$P$ to all densities will thus enter all the other transport coefficients.
This is why we examine the equation of state at {\em all densities} and
propose a {\em global equation of state} which is then tested by a
comparison with simulation and experiment.

For the numerical modeling of the system an event driven (ED) method
\cite{lubachevsky91,luding98f} is used.
A change in velocity can occur only at a collision when
the standard interaction model, based on momentum and energy
conservation is used \cite{herrmann98}.
The post-collisional velocities $\bv'$ of two collision partners,
in their center of mass reference frame, are given in terms of
the pre-collisional velocities $\bv$, by
$\bv_{1,2}' = \bv_{1,2} \mp \left [ (\bv_1 - \bv_2) \cdot \nn \right ] \nn$,
with the unit vector $\nn$, pointing along the line connecting the
centers of the colliding particles.  
This model can also be extended to the more general case of
dissipative particles with rough surfaces
\cite{jenkins85b,herrmann98}.

The stress tensor inside a test-volume $V$ (whose isotropic part is the pressure 
$p$) has two contributions, one from the convectional transport of mass and thus 
momentum and the other due to collisions and the related momentum transport,
for details see Refs.\ \cite{luding98f,luding00} and references therein.  
The mean pressure is obtained from 
simulations with different volume fractions $\nu$ in the following.

The equation of state in the dense, ordered phase has been
calculated by means of a free volume theory
\cite{kirkwood50buehler51wood52,berryman83,tobochnik88}, that leads in 2D 
to the reduced pressure $P_{\rm fv}=c_0/(\nu_{\rm max}-\nu)-1$,
with $c_0 \approx 1.8137$ as obtained from our numerical data.
Based on the simulation results we propose the corrected high
density pressure
\begin{equation}
P_{\rm dense}=\frac{c_0}{\nu_{\rm max}-\nu}  h_3(\nu_{\rm max}-\nu) -1
\label{eq:pdense}
\end{equation}
where $h_3(x)$ is a fit-polynomial $\left [ 1+c_1 x + c_3 x^3 \right ]$
of order three, with $c_1=-0.04$ and $c_3=3.25$ \cite{foot1}.

What remains to be done is to merge the low density pressure
$P_4$ and the high density expression (\ref{eq:pdense}).  
To our knowledge, no theory exists, which connects
these two limiting regimes. For various approaches concerning the
melting and freezing transition see Refs.\ 
\cite{hoover68,berryman83,tobochnik88,torquato95velasco97,fernandez95,santos99ric}. 
Therefore, we propose the {\em global equation of state}
\begin{equation}
Q = P_4 + m(\nu) [P_{\rm dense} - P_4] ~,
\label{eq:pgeneral}
\end{equation}
with an empirical merging function
\begin{equation}
m(\nu) = \frac{1}{1+\exp\left [-(\nu-\nu_c)/m_0 \right ]} ~
\end{equation}
which selects $P_4$ for $\nu \ll \nu_c$ and $P_{\rm dense}$ for
$\nu \gg \nu_c$, with the width of the transition $m_0$.
The parameters $\nu_c=0.7006$ and $m_0=0.0111$
lead to qualitative agreement between $Q$ and the
simulation results -- by far better than 1 per-cent for the purely
ordered and disordered regimes, and still within about 3 per-cent
in the interval $0.65 \le \nu \le 0.75$ \cite{foot2}.  

\FIGA
When plotting $P$ against the volume fraction $\nu$ with a logarithmic
vertical axis in Fig.\ \ref{fig:figQ}, the results for the different
simulations can not be distinguished from the theoretical
prediction $P_4$ for $\nu \stackrel{<}{\sim} 0.65$.  
For larger volume fractions one obtains crystallisation around 
$\nu_c \approx 0.70$ and the data clearly
deviate from $P_4$. In the transition regime, the coexistence of 
fluid and solid phases can be obtained. The pressure is strongly 
reduced due to the enhanced free volume in the ordered phase.
The reduced pressure data eventually diverge at the maximum packing
fraction in 2D $\nu^{\rm mono}_{\rm max}=\pi/(2\sqrt{3})$.
Note that one has to choose the system size such that a triangular
lattice fits perfectly in the system,
i.e.~$l_y/l_x=\sqrt{3} h/2w$ with integer $w$ and even $h$
 -- otherwise the maximum possible volume fraction can be smaller. Since 
our simulations are already set up on a perfect triangular lattice, the 
maximum density is approached for $\nu \rightarrow \nu_{\rm max}$.
If, instead, the volume fraction is increased by increasing the particle 
size \cite{lubachevsky91,torquato00} and the simulations are started from 
random, low density configurations, $\nu_{\rm max}$ is not 
reached due to defects in the 2D crystal \cite{hoover68,berryman83}.
Thus our global equation of state represents the high density, small
compression-rate limit.

The transition by itself is also interesting, since we obtain a hysteresis 
loop when the density is increased and decreased with a finite rate
\cite{alder55alder59alder62,fernandez95,luding98f}.
Especially in the transition region, the relaxation time is very large,
and the inflection in the data (see the inset in Fig.\ \ref{fig:figQ}) 
can be due to either the finite relaxation time, the finiteness of the 
system or the initial and boundary conditions \cite{fernandez95,foot2}.
Note that the analytical expression $Q$ allows for a straightforward
numerical integration of the density profile (see below), 
since the fit-parameters are chosen 
such that the slope of $Q$ is always positive.  
This is a compromise between the quality of the fit on the one hand and 
the numerical treatability of the function on the other hand -- instabilities
are avoided but also memory effects are disregarded. 

For a HS system in a gravitational field with the acceleration 
$g$ in the negative vertical direction, both density- and pressure
gradient have to be taken into account.  In the following, we compute
analytically the density profile for an ideal gas 
($\nu < 0.65$); the profile for the extremely dense gas is computed numerically
using the global equation of state and is found to be in excellent agreement
with the numerical ED simulations, where a horizontal wall at $z=0$ is 
introduced in a periodic, two-dimensional system of width $L=l_x/(2a)$ 
and infinite height. The number density $n=n(z)=N/V$ is related to the
volume fraction by $n=\nu(z)/(\pi a^2)$. Here, we briefly sketch 
how to obtain an analytical solution for the density profile,
valid at least for low and intermediate densities 
\cite{sunthar99,hong99,luding00}.

Given the equilibrium of forces, the force $-L{d}p$ due to the pressure 
gradient at height $z$ compensates the weight $n m g L {d}z$ of the particles 
in a layer with height $dz$, so that the differential equation ${dp}/{dz} = 
-n m g$ has to be solved. In the simplest case, the equation of state of an 
ideal dilute gas $p=nT$, separation of variables, and the assumption of a constant 
temperature, leads to an exponentially decreasing density profile
$\nu(z) = \nu_d \exp \left( -(z-z_0)/z_T \right)$,
with $\nu<\nu_d:=\nu(z_0)$ and $z_T=T/(mg)$. 
In a closed system, the particle number $N$ is conserved so that
integration of $n$ over $z$ determines the volume fraction at the 
bottom $\nu_d=N \pi a^2/(z_T L)$, in the dilute limit.

In denser situations ($0 < \nu < 0.65$) the pressure can be expressed
as $p=nT(1+2 \nu g_2(\nu))$ [we do not use $g_4(\nu)$ in order to keep 
the analysis simple], and integration leads to an implicit definition 
of $\nu(z)$:
\begin{equation}
  \label{eq:znu}
  \frac{z-z_0}{z_T} = 
    \ln\frac{\nu_0}{\nu}-\frac{7}{8}\ln\frac{1-\nu_0}{1-\nu}+
     2 [g_2(\nu_0) - g_2(\nu)]
\end{equation}
with the unknown volume fraction $\nu_0$ at $z_0$, again
determined by the integral over the density. 
This leads to a third order polynomial for $\nu_0$, 
which can be solved analytically \cite{bronstein79}, and has 
at least one real solution. When the theoretical density profile 
in Eq.\ (\ref{eq:znu}) is compared with numerical simulations,
one obtains perfect agreement for $\nu < 0.65$ \cite{luding00}.
Since the functions $g_2(\nu)$ and $g_4(\nu)$ are wrong 
at larger densities $\nu$, one cannot expect that the 
pressure and the density profile are correct.  

Using the gobal equation of state, $Q$, from Eq.\ (\ref{eq:pgeneral}) 
instead of $2\nu g_{2}(\nu)$, one has to integrate
the differential equation $dp/dz=- n m g$ numerically with 
$p=nT(1+Q)$ under the constraint that the particle number is a constant.
In Fig.\ \ref{fig:grav1}, the volume fraction $\nu$ is plotted against 
the rescaled height $z/z_T$ for both theory and simulations.
Simulation parameters are $N=1000$, $L=10$, and $z_T/(2a)=5.85$ (open 
circles) or $N=3000$, $L=50$, and $z_T/(2a)=0.508$ (open squares).  
In addition, we present experimental results from vibrated two-dimensional 
arrays of small spheres \cite{clement91} (solid dots), neglecting the fact 
that this situation is weakly dissipative.
Both the qualitative and the quantitative behavior of the density profile
is well reproduced by the numerical solution using the global equation
of state. All solutions belong to one master curve and can be rescaled
by a horizontal shift.  The averaging result is 
somewhat dependent on the averaging procedure -- we evidence
strong coarse-graining effects in the dense, ordered regime with 
densities $\nu > 0.70$. Using two methods, one tailord for the ordered
regime and the other for the disordered regime, however, we obtain
consistent results.  
\FIGB

In summary, we tested existing predictions for the equation of state of a 
2D hard sphere gas of arbitrary density via comparison with numerical 
simulations and experimental data. 
In the dilute case, the particle correlation at contact and the collision 
frequency (and thus the equation of state) are nicely predicted by the 
kinetic theory expressions up to intermediate densities
$\nu\approx0.65$.  In the dense case, the free volume theory for 2D systems 
can be applied to systems with densities larger than $\nu\approx0.75$.
Finally, a merging function is proposed, which connects the low and 
high density regimes, resulting in a differentiable {\em global equation 
of state} for the 2D hard sphere gas for arbitrary density.

The equation of state is used to compute analytically and numerically 
the density profile of an elastic, monodisperse HS gas in a
gravitational field.  For maximum densities below 
$\nu_c$, the analytical solution works perfectly well, 
for higher densities, we used a numerical solver (MAPLE). The
strange shape of the density profile as obtained from simulations 
is nicely reproduced by our theory based on the global equation of state, 
including a wiggle at $\nu\approx\nu_c$.
We remark that the ED simulation method parallels the Monte Carlo (MC) 
method \cite{fernandez95} concerning the particle-particle interactions, 
but in contrast to MC allows for a definition of time and thus for the 
examination of the dynamics.

The presented results are obtained from homogeneous,
elastic systems of arbitrary density. The range of applicability,
however, is much wider. Since already weak dissipation can lead to strong 
inhomogeneities in density, temperature, and pressure, the global equation 
of state is a necessary tool to treat effects like clustering, surface-waves, 
pattern formation, or phase-transition and -coexistence by means of a continuum 
theory.  In a freely cooling ``granular gas'', for example, clustering leads to
{\em all} densities between $\nu \approx 0$ and $\nu \approx \nu_{\rm max}$
\cite{luding98f}.

The proposed global equation of state is based on a limited amount of data from 
ED simulations. First checks, whether our global equation of state still makes 
sense for different particle-size distribution functions are promising,
however, the crystallisation effect vanishes for strong enough polydispersity
\cite{luding00}.
What remains to be done is to find similar {\em global} expressions for 
other transport coefficients like the viscosity and the heat-conductivity 
and, furthermore, to extend the presented approach to three dimensional 
systems. \\
We acknowledge the support of the Deutsche Forschungsgemeinschaft 
(DFG) and appreciate the helpful discussions with E. Cl\'ement, 
J. Eggers, D. Hong, J. Jenkins, A. Santos, and O. Strau\ss{}.


\begin{thebibliography}{10}
\bibitem{multicont}
The instantaneous contacts of the particles imply that exclusively
two particle interactions take place.  Multi-body interactions can be 
accounted for via higher order correlations \cite{landau86,noije98c} 
or by perturbations to the rigid potential, see section 6.3. in Ref.\
\cite{hansen86}.
\bibitem{foot1}
The magnitude of the correction at $\nu=0.75$ is only $0.5$ per-cent so 
that one could also use $c_1=c_3=0$ with reasonable accuracy so that 
$P_{\rm dense}=P_{\rm fv}$.
\bibitem{foot2}
The best-fit parameters $m_0=0.00688$ and $\nu_c=0.7022$ lead to even
better agreement but would reproduce the inflection (negative slope) of
$Q$ within about 1.5 per-cent, in the transition regime $0.65<\nu<0.75$.
\end{thebibliography}

~\vspace{-1.2cm}\\

\end{document}